\documentclass[fleqn,usenatbib]{mnras}

\usepackage{newtxtext,newtxmath} 
\usepackage{graphicx}
\usepackage{url}
\usepackage{multirow}

\title[Companion masses of Gaia DR3 AstroSpectroSB1 binaries]{Companion masses of Gaia DR3 AstroSpectroSB1 binaries: a Bayesian forward-modelling framework}

\author[Prasid Sandave and Raghuveer Garani]{
Prasid Sandave\thanks{E-mail: ph21b007@smail.iitm.ac.in}
and Raghuveer Garani\thanks{E-mail:garani@iitm.ac.in}
\\
Centre for Strings, Gravitation and Cosmology, Department of Physics, Indian Institute of Technology Madras, Chennai 600036, India
}

\date{}
\makeatletter
\journal{}
\let\@oddfoot\@empty
\let\@evenfoot\@empty
\makeatother

\begin{document}
\label{firstpage}
\pagerange{\pageref{firstpage}--\pageref{lastpage}}
\maketitle

\begin{abstract}
Gaia Data Release 3 (DR3) has substantially expanded the study of unresolved binaries by providing combined astrometric and spectroscopic orbital solutions for a large sample of non-single stars. We present a Bayesian forward-modelling framework for estimating companion masses in Gaia DR3 \texttt{AstroSpectroSB1} systems. The method samples the physical parameters of each binary, predicts the corresponding astrometric Thiele-Innes constants and spectroscopic orbital coefficients, and evaluates their likelihood using the full covariance matrix published by Gaia. An external prior on the primary mass is incorporated to resolve the remaining degeneracy between the component masses and orbital inclination in single-lined systems. We apply this framework to 1035 \texttt{AstroSpectroSB1} systems with primary-mass estimates drawn from the Gaia DR3 \texttt{astrophysical\_parameters} table. Of these, 760 systems also have Gaia-reported estimates for both component masses and are used as a validation sample. The remaining 275 systems constitute an inference sample for which the mass of the fainter, unresolved companion is estimated. For systems with well converged posteriors, the inferred parameters reproduce the Gaia orbital observables, while the companion masses recovered for the validation sample are consistent with the Gaia estimates. In the inference sample we find 113 well converged systems for which the companion masses are meaningfully inferred. We find that the population is dominated by ordinary stars, together with a smaller high-mass tail that may contain massive white dwarfs or other compact-object candidates. Our framework provides a scalable approach for converting Gaia \texttt{AstroSpectroSB1} solutions into companion-mass posterior distributions, identifying promising systems for targeted follow-up, and preparing for the larger and more precise orbital samples anticipated with Gaia DR4.
\end{abstract}


\section{Introduction}
Binary stars are fundamental laboratories for stellar astrophysics. They provide some of the most direct constraints on stellar masses, calibrate models of stellar structure and evolution, and form the progenitors of many compact-object and transient systems, including X-ray binaries, Type Ia supernovae, millisecond pulsars, and gravitational-wave sources. Yet many binaries remain difficult to characterize, especially when the system is unresolved and only one component contributes appreciably to the observed light. In such systems the measured astrometric motion traces the photocentre rather than the relative orbit, and the physical interpretation of the orbit depends on combining astrometric, spectroscopic, and photometric information.

The Gaia mission has transformed this problem by providing high-precision astrometry for more than a billion Galactic sources. While the first Gaia data releases primarily advanced single-star astrometry, Gaia Data Release 3 (DR3) marked the first large-scale release of non-single-star (NSS) solutions. These include astrometric, spectroscopic, eclipsing, acceleration, and joint astrometric--spectroscopic orbital solutions, thereby opening a new statistical window onto unresolved binaries across a wide range of periods and companion masses \citep{Gaia:2016, Gaiadr3:2023, Halbwachs:2023A&A, Holl:2023A&A, 2024NewAR..9801694E}. The Gaia DR3 multiplicity catalogue increased the number of available binary orbital solutions by more than an order of magnitude and demonstrated the broad scientific potential of these systems, from stellar binaries and brown-dwarf companions to white dwarfs, neutron-star candidates, and black-hole candidates \citep{Arenou:2023A&A}.

Of particular interest are systems in which the unseen companion may be a compact object. Dormant compact-object binaries with luminous companions are expected from binary evolution, but they are observationally difficult to identify because they need not emit X-rays or show accretion signatures. Astrometry is especially well suited to this task: a dark massive companion can produce a detectable photocentre wobble even when it is electromagnetically invisible or partially visible. Several works have therefore used Gaia DR3 astrometric binaries to search for compact companions, often through an astrometric mass-ratio function or related triage criteria designed to test whether the observed orbit can be produced by a normal luminous companion \citep{Shahaf:2019MNRAS, shahaf:2023, Arenou:2023A&A, Yamaguchi:2024MNRAS, Shahaf:2024MNRAS}.  The discovery and follow-up of Gaia black-hole and neutron-star candidates have emphasized both the power and the difficulty of this approach. Gaia BH systems have demonstrated that astrometry can uncover dormant stellar-mass black holes in au-scale orbits \citep{ElBadry:2023MNRASBH, ElBadry:2023MNRASRG, Chakrabarti:2023AJ, Panuzzo:2024A&A}. At lower companion masses, the interpretation is more subtle: neutron-star candidates often lie close to the maximum white-dwarf mass, so massive white dwarfs or unresolved compact inner binaries can mimic a neutron-star companion unless the dynamical mass is tightly constrained. Recent follow-up studies have therefore combined Gaia astrometry with multi-epoch radial velocities, stellar-parameter estimates, abundance measurements, and checks for contaminating light to identify and characterize candidate neutron-star companions \citep{ElBadry:2024OJApNS1, ElBadry:2024OJApNS2}. These studies show that the mass inferred for the unseen companion can depend sensitively on the assumed primary mass, metallicity, flux ratio, inclination, and possible systematics in the Gaia orbital solution.
Complementary searches have used other Gaia data products to identify compact-object candidates. Short-period ellipsoidal variables provide a photometric route to massive unseen companions, since tidal distortion of a luminous star can produce periodic variability even when the companion is dark~\citep{Gomel:2023A&A, 2024NewAR..9801694E}. Population-synthesis studies have also begun to quantify what Gaia DR3 should have detected and what future Gaia releases may reveal. These works suggest that DR3 is only the beginning: the end-of-mission Gaia sample should contain substantially larger populations of white-dwarf, neutron-star, and black-hole binaries with luminous companions~\citep{Chawla:2025}. A robust, scalable method for propagating Gaia orbital uncertainties into companion-mass posteriors is therefore timely.

In this work, we focus on Gaia DR3 systems with \texttt{AstroSpectroSB1} solutions \citep{Halbwachs:2023A&A}. These binaries are particularly informative because Gaia provides both sky-plane astrometric information and line-of-sight spectroscopic constraints. The orbital solution is reported through astrometric Thiele-Innes coefficients together with spectroscopic coefficients, and the full covariance matrix encodes important correlations among the fitted parameters. In principle, this information is sufficient to connect the observed photocentre orbit to the underlying physical binary parameters, including the companion mass. In practice, however, the companion mass may not be a direct Gaia catalogue observable, and a statistically consistent inference requires propagating the full covariance structure together with information on the luminous primary.
We develop a Bayesian forward-modelling framework to infer companion masses for Gaia DR3 \texttt{AstroSpectroSB1} binaries. The model samples the physical binary parameters, maps them to the Gaia-observed astrometric and spectroscopic coefficients, and evaluates the likelihood using the published covariance matrix.  External stellar information from Gaia table ~\texttt{astrophysical\_parameters} is used to place priors on the luminous primary. We validate the method on a subset of systems for which both component masses are independently available, and then apply it to a larger sample in which the companion mass is inferred. This approach is complementary to compact-object triage studies: rather than applying a deterministic classification criterion, we produce posterior distributions for companion masses and identify systems whose inferred masses make them promising targets for follow-up.

The paper is organized as follows. Section~\ref{sec:data} describes the construction of the Gaia DR3 \texttt{AstroSpectroSB1} sample analysed in this paper, including the external wide-binary catalogue used as an independent multiplicity check and the Gaia mass information that supplies the prior on the more luminous component of the binary system. Section~\ref{sec:methods} presents the forward model, likelihood, priors, and MCMC sampling procedure used to infer companion masses. In Section~\ref{sec:results}, we present the inferred companion-mass distribution for the inference sample, validate the framework against systems with independently known companion masses, and discuss the dependence of the constraints on orbital geometry. Section~\ref{sec:conclusion} summarizes our conclusions and discusses possible limitations and selection effects in our analysis. Further details on the MCMC implementation and the validation exercise are given in Appendices~\ref{app:mcmc_details} and~\ref{app:validation}, respectively.

\section{Data selection}
\label{sec:data}
We identify the sample of Gaia DR3 \textsc{AstroSpectroSB1} systems analysed in this paper, and obtain the mass information needed to interpret them, in three steps. (1) We construct an external, independent catalogue of wide binaries across the whole of Gaia DR3, following~\citep{million_binaries}, see Section~\ref{sec:widecat}. This catalogue is built without reference to Gaia's non-single-star pipeline, and does not by itself supply any mass used in this paper. (2) We cross-match this catalogue against the Gaia DR3 \textsc{AstroSpectroSB1} solutions and retain only the intersection: hosts independently confirmed to be part of a multiple-star system by this second, NSS-independent method (Section~\ref{sec:crossmatch}). (3) For this intersection sample, we obtain primary-mass information from Gaia's \texttt{astrophysical\_parameters} table, see Section~\ref{sec:binarymasses}. Throughout, $M_1$ and $M_2$ denote the two components of the \emph{close}, \textsc{AstroSpectroSB1} solutions, i.e. the luminous primary and its fainter companion, respectively.

\subsection{External wide-binary catalogue}
\label{sec:widecat}
We construct a catalogue of wide binaries across Gaia DR3, following the general strategy of~\citep{million_binaries}, updated here to DR3. We begin with all Gaia DR3 sources satisfying
\begin{itemize}
  \item $\varpi > 1$~mas (restricting the sample to distances within approximately 1~kpc),
  \item $\varpi/\sigma_\varpi > 5$,
  \item $\sigma_\varpi < 2$~mas,
  \item an available $G$-band magnitude,
\end{itemize}
which returns $N = 64\,407\,858$ candidate sources. Following Section~2 of~\cite{million_binaries}, we impose three criteria on candidate pairs drawn from this set. First, a projected-separation cut,
\begin{equation}
  \theta \le 206.265\ \frac{\varpi_{\rm A}}{\rm mas}\ {\rm arcsec},
  \label{eq:sepcut}
\end{equation}
where $\theta$ is the angular separation of the pair and $\varpi_{\rm A}$ is the parallax of its brighter component. This corresponds to a maximum physical separation of 1~pc. Near this separation the Galactic tidal field becomes comparable to the pair's mutual gravitational attraction, beyond which essentially no bound pairs are expected to survive \citep{galatic_dynamic_book}. A binary at this separation has an orbital period of order $10^{8}$~yr -- roughly eight orders of magnitude longer than the days-to-years periods spanned by the \textsc{AstroSpectroSB1} sample itself. We return to this scale separation in Section~\ref{sec:crossmatch}, where it is the reason a match between the two catalogues can be interpreted unambiguously.
Second, a parallax-consistency cut,
\begin{equation}
  |\varpi_{\rm A} - \varpi_{\rm B}| < b\sqrt{\sigma_{\varpi,{\rm A}}^2 + \sigma_{\varpi,{\rm B}}^2},
  \label{eq:plxcut}
\end{equation}
with $b = 6$ for $\theta < 4$~arcsec and $b = 3$ for $\theta \ge 4$~arcsec. The looser threshold at small separation reflects two effects that both push in the same direction there: the chance-alignment rate is lower at small separation (so a looser cut costs little in purity), while parallax uncertainties are known to be underestimated for close pairs \citep{million_binaries} (so a strict cut would reject genuine pairs). Third, a proper-motion-consistency cut,
\begin{eqnarray}
  \Delta\mu &\le& \Delta\mu_{\rm orbit} + 2\sigma_{\Delta\mu}, \nonumber \\
  \Delta\mu_{\rm orbit} &=& 0.44\ {\rm mas\,yr^{-1}}\left(\frac{\varpi}{\rm mas}\right)^{3/2}\left(\frac{\theta}{\rm arcsec}\right)^{-1/2}~.
  \label{eq:pmcut}
\end{eqnarray}
Here $\Delta\mu$ is the observed scalar proper-motion difference between the two components and $\sigma_{\Delta\mu}$ its uncertainty and $\Delta\mu_{\rm orbit}$ is the maximum proper-motion difference attributable to the pair's own orbital motion at the separation implied by $\theta$ and $\varpi$, i.e. two stars sharing a common space motion are not expected to have \emph{identical} proper motions, only proper motions consistent with orbiting a common barycentre.

We then remove chance alignments, dense-field contamination, and higher-order multiples in three further steps: (i) using BallTree spatial index (Haversine metric) flags, and we reject, any star with more than 30 neighbours within a 5~pc radius of consistent parallax ($2\sigma$) and proper motion (within 5~km\,s$^{-1}$). Thus removing the majority of spurious pairs in dense fields. (ii) We remove any pair in which either component's source ID recurs in another candidate pair, excluding resolved triples and residual chance superpositions from this catalogue. (iii) We reject any remaining candidate with more than one neighbouring pair under the same criteria, therefore removing small clusters or comoving groups not caught by (i). This leaves a final catalogue of $1\,815\,389$ wide pairs: 2.82~per cent of the $64\,407\,858$ candidate sources counted as sources, or equivalently 5.64~per cent counted as individual stars ($3\,630\,778 = 2\times 1\,815\,389$). The resulting Hertzsprung--Russell diagram reproduces the features reported by~\cite{million_binaries}. The median parallax of the full catalogue is $1.8845$~mas, and the median $G$ magnitude is $15.6199$ for the brighter and $18.2359$ for the fainter component.
We do not extract or use a component mass from this catalogue at any point: no such step is part of our pipeline, for either component of a wide pair. Its role in what follows (Section~\ref{sec:crossmatch}) is solely to establish, independently of Gaia's non-single-star pipeline, that a given \textsc{AstroSpectroSB1} host has a resolved companion at wide separation and not to constrain $M_1$ or $M_2$. 

\subsection{Cross-match with Gaia DR3 AstroSpectroSB1 solutions}
\label{sec:crossmatch}
We cross-match the $3\,630\,778$ individual stars of the wide-binary catalogue against the Gaia DR3 \texttt{nss\_two\_body\_orbit} table \citep{Halbwachs:2023A&A, Holl:2023A&A, Gosset:2025A&A, KatzRV:2023}, retaining matches with solution type \textsc{AstroSpectroSB1}. Of the $33467$ \textsc{AstroSpectroSB1} solutions in Gaia DR3, $1665$ are hosted by a star that also appears in the wide-binary catalogue, as either component of its pair.  This requirement restricts the sample to \textsc{AstroSpectroSB1} hosts that are demonstrably part of a hierarchical, close-plus-wide multiple-star system entirely independent of Gaia's non-single-star pipeline. We therefore restrict our analysis to this intersection: the $1665$ \textsc{AstroSpectroSB1} systems with an independently confirmed companion are the sample analysed in the remainder of this paper, see also Tab.~\ref{tab:samplesizes}.

\begin{table}
  \centering
  \caption{Parent-sample sizes and pairwise/three-way intersections among the three Gaia DR3 products used in Section~\ref{sec:data}: the wide-binary catalogue (WBC, Section~\ref{sec:widecat}), the \textsc{AstroSpectroSB1} solutions in \texttt{nss\_two\_body\_orbit} (SB1), and \texttt{binary\_masses} (BM, Section~\ref{sec:binarymasses}). $N_{\rm match}$ in the text of Section~\ref{sec:crossmatch} is the WBC$\,\cap\,$SB1 row; $N_{\rm bm}$ is the BM$\,\cap\,$SB1$\,\cap\,$WBC row. $N_{\rm bm}$ splits into the validation and inference samples analysed in Section~\ref{sec:results} onward, according to whether \texttt{binary\_masses} also reports $M_2$.}
  \label{tab:samplesizes}
  \small
  \begin{tabular}{lr}
    \hline
    Quantity & $N$ \\
    \hline
    Wide-binary catalogue (WBC)                         & 1815389 \\
    \texttt{nss\_two\_body\_orbit} (all solution types)  & 443205 \\
    \textsc{AstroSpectroSB1} solutions in NSS (SB1)      & 33467 \\
    WBC $\cap$ SB1 ($N_{\rm match}$)                     & 1665 \\
    BM $\cap$ WBC                                        & 10362 \\
    BM $\cap$ SB1                                        & 20213 \\
    \hline
  \end{tabular}
\end{table}

\subsection{Mass information: \texttt{astrophysical\_parameters} and \texttt{binary\_masses}}
\label{sec:binarymasses}
The \texttt{nss\_two\_body\_orbit} table stores the astrometric Thiele--Innes coefficients, the spectroscopic coefficients, and the orbital elements (see Section~\ref{sec:methods} for details) for $443205$ objects. For solutions like \textsc{AstroSpectroSB1} these alone cannot separate $M_1$ from $M_2$ and orbital orientation (e.g. $\sin i$, see section~\ref{sec:methods}): an external constraint on the luminous primary is required to break this degeneracy. This is precisely the role $M_1$ plays as a prior in Section~\ref{sec:methods}.
We obtain this constraint from the Gaia DR3 \texttt{astrophysical\_parameters} table (AP), specifically its \texttt{mass\_flame} field: a FLAME-derived stellar mass for the host star, based on photometry, parallax, and derived luminosity and effective temperature, and independent of the \textsc{AstroSpectroSB1} orbital solution itself. As with the wide-binary cross-match, AP entries are indexed by the same \texttt{source\_id} as the \textsc{AstroSpectroSB1} host itself. No property of the wide companion identified in Section~\ref{sec:widecat} enters this step. Of the $1665$ \textsc{AstroSpectroSB1} systems with an independently confirmed wide companion (Section~\ref{sec:crossmatch}), $1035$ have an AP \texttt{mass\_flame} entry (Table~\ref{tab:massavail}). We adopt this value, together with its reported uncertainty, as $\mu_{M_1}$ and $\sigma_{M_1}$ in the Gaussian prior on the primary mass used in Section~\ref{sec:methods}.

For a subset of these $1035$ systems we additionally query the derived table \texttt{binary\_masses} (BM), described in \citep{Arenou:2023A&A, Halbwachs:2023A&A}, by \texttt{source\_id}. The table \texttt{binary\_masses} contains $195\,315$ rows across several combination methods keyed to NSS solution type (e.g.\ \texttt{Orbital+SB2}, \texttt{Eclipsing+SB2}, \texttt{AstroSpectroSB1+M1}, \texttt{SB1+M1}). For solutions such as \textsc{AstroSpectroSB1}, Gaia derives $M_1$ (recorded as \texttt{m1\_ref = `IsocLum'}) by placing the primary on a main-sequence isochrone using its photometry and parallax, then steps the assumed flux ratio in increments of $0.01$, and at each step computes the corresponding $M_2$ from the orbit and checks whether that $M_2$ is itself consistent with a main-sequence star at the tested flux ratio. Only self-consistent $(M_1, M_2, \text{flux ratio})$ combinations are retained. 

Solutions are handled differently, or excluded, in three specific cases: an evolved (non-main-sequence) primary, whose mass is degenerate with age and metallicity at fixed luminosity and is therefore not derived this way; a companion independently identified as a white dwarf (e.g.\ via cross-match against a Gaia-based white-dwarf catalogue such as \citep{Fusillo_WD_cat}), which is instead assigned a fixed mass of $0.65 \pm 0.16\,M_\odot$; and an astrometric or spectroscopic signal-to-noise below 5, in which case the solution is treated as \texttt{Orbital}-only and receives no \texttt{binary\_masses} entry. We use only BM's $M_2$ (\texttt{m2}) here, as an independent check on our own inferred companion mass; BM's $M_1$ plays no role in our analysis, since the AP \texttt{mass\_flame} value is adopted throughout. Where BM reports \texttt{m2} for a system with an AP \texttt{mass\_flame} entry, that system enters the validation sample of Section~\ref{sec:methods}: $760$ of the $1035$ systems qualify on this basis. The remaining $275$ systems, for which AP reports only $M_1$ and BM reports no companion mass, form the inference sample, for which $M_2$ is the quantity inferred in Section~\ref{sec:methods}. We summarize this information in Table~\ref{tab:samplesizes}. In Table~\ref{tab:massavail} we document the underlying AP \texttt{mass\_flame} $\cap$ \texttt{binary\_masses} availability within the full $1665$-system sample.

\begin{table}
  \centering
  \caption{Breakdown of the $N_{\rm match}=1665$ WBC$\,\cap\,$SB1 sample of Table~\ref{tab:samplesizes} by availability of the AP \texttt{mass\_flame} field (host or primary mass) and the \texttt{m2} field (companion mass), and the resulting sample assignment used from Section~\ref{sec:results} onward. The $604$ row recovers systems for which \texttt{m2} was not returned by the initial WBC$\,\cap\,$SB1 cross-match but is present in \texttt{binary\_masses} (BM) under a direct \texttt{source\_id} query. These systems correspond to the validation sample, as described in Section~\ref{sec:binarymasses}.}
  \label{tab:massavail}
  \begin{tabular}{llrl}
    \hline
    AP \texttt{mass\_flame} &  \texttt{m2} status & $N$ & Sample \\
    \hline
    \multirow{3}{*}{available (1\,035)}
      & available         & 156 & \multirow{4}{*}{Validation (760)} \\
      &   (cross-match)     &   & \\
      & recovered via   & 604 & \\
      &  direct BM query                   &               &\\
      \cline{2-4}
      & unavailable                    & 275 & Inference (275) \\
    \cline{1-4}
                &                   &   &           \\
    not reported  & --            & 630 & Excluded \\
        (630)                &               &       &\\
    \hline
    Total               &              & 1\,665 & \\
    \hline
  \end{tabular}
\end{table}

\section{Methods and inference}
\label{sec:methods}
\subsection{Data vector}
\label{subsec:data_vector}
For each \texttt{AstroSpectroSB1} source, we define the Gaia data vector as
\begin{equation}
\mathbf{d} = \left( \alpha,\delta,\varpi,\mu_{\alpha *},\mu_\delta, A,B,F,G,C,H,\gamma,e,P,T_0 \right),
\label{eq:data_vector}
\end{equation}
where $\alpha$ and $\delta$ are the sky coordinates, $\varpi$ is the parallax, $\mu_{\alpha *}$ and $\mu_\delta$ are the proper motions, $A,B,F,G$ are the astrometric Thiele--Innes coefficients, $C,H$ are the spectroscopic orbital coefficients, $\gamma$ is the centre-of-mass velocity, $e$ is the eccentricity, $P$ is the orbital period, and $T_0$ is the time of periastron passage \citep{Arenou:2023A&A, Gaiacollab:2023A&Anss}. The covariance matrix is constructed from the Gaia-reported parameter uncertainties and correlation coefficients,
\begin{equation}
\Sigma_{ij} = \rho_{ij}\sigma_i\sigma_j,
\label{eq:covariance}
\end{equation}
where $\sigma_i$ is the uncertainty on the $i$th entry of the data vector and $\rho_{ij}$ is the corresponding correlation coefficient. Since the data vector has fifteen entries, the full covariance matrix contains 105 independent off-diagonal correlations. We retain only systems for which the covariance information required for the likelihood evaluation is available. Before evaluating the likelihood, the covariance matrix is symmetrized and checked for positive definiteness. If small numerical eigenvalues prevent a stable factorization, we add a diagonal regularization term,
\begin{equation}
\Sigma \rightarrow \Sigma + \lambda I,
\end{equation}
where $\lambda$ is chosen to be the minimum value required to make the matrix numerically positive definite. The likelihood is then evaluated using a Cholesky decomposition whenever possible. If the Cholesky decomposition fails, we use a singular-value-decomposition-stabilized inverse.

\subsection{Sampled parameters}
\label{subsec:sampled_parameters}
The MCMC is performed in terms of physical and geometric binary parameters rather than directly in terms of the Gaia Thiele--Innes coefficients. The sampled parameter vector is

\begin{multline}
\boldsymbol{\theta}= \big( \alpha,\delta,\varpi,\mu_{\alpha *},\mu_\delta,
P,e\cos\omega,e\sin\omega,T_0,\gamma,M_1,M_2,\cos i,\Omega\big).
\label{eq:theta_vector}
\end{multline}

Here $M_1$ is the mass of the luminous primary, $M_2$ is the companion mass, $i$ is the orbital inclination, $\Omega$ is the longitude of the ascending node, and $\omega$ is the argument of periastron.
We sample in the variables $e\cos\omega$ and $e\sin\omega$ rather than in $e$ and $\omega$ directly. This choice avoids the coordinate singularity at $e=0$ and improves numerical sampling for nearly circular orbits. At each likelihood evaluation, we reconstruct

\begin{equation}
e=\sqrt{(e\cos\omega)^2+(e\sin\omega)^2},
\label{eq:e_reconstruct}
\end{equation}
and
\begin{equation}
\omega=\arctan2(e\sin\omega,e\cos\omega).
\label{eq:omega_reconstruct}
\end{equation}

The inclination is sampled through $\cos i$, with $\cos i\in[-1,1]$, corresponding to an isotropic prior on orbital orientation.

\subsection{Forward model}
\label{subsec:forward_model}
For each proposed parameter vector $\boldsymbol{\theta}$, we compute a model prediction for the Gaia data vector,
\begin{multline}
\mathbf{m}(\boldsymbol{\theta})=\big(
\alpha,\delta,\varpi,\mu_{\alpha *},\mu_\delta,A,B,F,G,C,H,\gamma,e,P,T_0\big)_{\rm model}.
\label{eq:model_vector}
\end{multline}

The astrometric parameters, $P$, $T_0$, and $\gamma$ are sampled directly, while $e$ and $\omega$ are reconstructed from Eq.~\eqref{eq:e_reconstruct} and Eq.~\eqref{eq:omega_reconstruct}. The Thiele--Innes and spectroscopic coefficients are then computed from the physical binary model. The relative semi-major axis is obtained from Kepler's third law,
\begin{equation}
a_{\rm rel} = \left(M_1+M_2\right)^{1/3} \left(\frac{P}{365.25}\right)^{2/3}~,
\label{eq:arel}
\end{equation}
where $a_{\rm rel}$ is in AU, $P$ is in days, and the masses are in solar masses. The semi-major axis of the luminous primary about the barycentre is
\begin{equation}
a_1= a_{\rm rel} \frac{M_2}{M_1+M_2}.
\label{eq:a1}
\end{equation}

Gaia measures the orbit of the photocentre rather than the relative orbit of the two components. The secondary-to-primary flux ratio is defined as
\begin{equation}
\epsilon = \frac{F_2}{F_1}~,
\end{equation}
and the corresponding fractional light contribution of the secondary is
\begin{equation}
f=\frac{\epsilon}{1+\epsilon}~.
\end{equation}

In the present analysis, $\epsilon$ (and hence $f$) is fixed using the available Gaia component G-band flux information where available. This treatment keeps the photometric contribution fixed while allowing the dynamical quantities to vary in the MCMC. The angular semi-major axis of the photocentre is then expressed as
\begin{equation}\label{eq:a0}
a_0 = \varpi\,a_{\rm rel} \left[ \frac{M_2}{M_1+M_2}-f \right].
\end{equation}

Here $\varpi$ is expressed in arcseconds while $a_{\rm rel}$ is expressed in AU. The astrometric Thiele--Innes coefficients are then computed as follows:
\begin{align}
A &= a_0 \left(\cos\omega\cos\Omega - \sin\omega\sin\Omega\cos i \right), \\
B &= a_0 \left( \cos\omega\sin\Omega + \sin\omega\cos\Omega\cos i\right), \\
F &= a_0 \left(-\sin\omega\cos\Omega - \cos\omega\sin\Omega\cos i \right), \\
G &= a_0 \left( -\sin\omega\sin\Omega + \cos\omega\cos\Omega\cos i \right)~.
\end{align}
The spectroscopic coefficients are computed from the same physical orbit as
\begin{align}
C &= a_1 \sin\omega \sin i~,\\
H &= a_1 \cos\omega \sin i~.
\end{align}

Thus, the astrometric and spectroscopic parts of the Gaia solution are not fitted independently, both are generated by the same underlying masses and orbital geometry.

\subsection{Likelihood}
\label{subsec:likelihood}
The likelihood compares the Gaia data vector $\mathbf{d}$ with the model prediction $\mathbf{m}(\boldsymbol{\theta})$. Assuming Gaussian errors in the Gaia-published parameters, we use
\begin{equation}
\ln \mathcal{L}=-\frac{1}{2}\Delta\mathbf{d}^{T}\Sigma^{-1}\Delta\mathbf{d}~,
\label{eq:likelihood}
\end{equation}
where $\Delta\mathbf{d} =\mathbf{d}-\mathbf{m}(\boldsymbol{\theta})$. The normalization term involving $\det\Sigma$ is independent of the model parameters for a given source and is therefore omitted in the MCMC sampling. In practice, we do not explicitly compute $\Sigma^{-1}$ when a stable factorization is available; instead, the quadratic form is evaluated using the Cholesky or SVD representation of the covariance matrix.

\subsection{Priors}
\label{subsec:priors}
The primary mass is assigned a Gaussian prior based on the Gaia-provided mass estimate (see Section~\ref{sec:binarymasses}),
\begin{equation}
M_1 \sim \mathcal{N} \left( \mu_{M_1}, \sigma_{M_1}\right)~.
\label{eq:M1_prior}
\end{equation}
When Gaia reports asymmetric upper and lower uncertainties, we use the effective symmetric uncertainty
\begin{equation}
\sigma_{M_1} = \frac{M_{1,\rm upper}-M_{1,\rm lower}}{2}.
\end{equation}
The companion mass is assigned a uniform prior with hard bounds,
\begin{equation}
M_2 \sim U(0.1,3.0)\,M_\odot,
\end{equation}

the same range used to initialize the walkers. This range is broad enough that, for well-converged systems, the posterior is determined by the Gaia likelihood rather than by the prior. It is not, however, unbounded: a hard ceiling at $3.0\,M_\odot$ means the prior itself forbids any companion mass above this value, so this analysis cannot by construction distinguish a companion at the prior edge from one whose true mass lies higher still. Systems whose posterior mass piles up against $M_2=3.0\,M_\odot$ should be flagged as prior-bounded rather than data-constrained. For the orbital orientation we use
\begin{equation}
\cos i \sim U(-1,1), \qquad \Omega \sim U(0,360^\circ).
\end{equation}

The eccentricity is restricted to the physical range
\begin{equation}
0 \leq e < 0.999~,
\end{equation}
and we require $\varpi>0, \, P>0$~. The systemic velocity is assigned a broad uniform prior,
\begin{equation}
\gamma \sim U(-1000,1000)\ {\rm km\,s^{-1}}~.
\end{equation}
Because the sampling variables are $e\cos\omega$ and $e\sin\omega$, we include the appropriate Jacobian factor so that the induced prior is flat in $e$ rather than flat in the Cartesian eccentricity components.

\subsection{MCMC implementation}
\label{subsec:mcmc}
We sample the posterior distribution
\begin{equation}
p(\boldsymbol{\theta}|\mathbf{d})
\propto
\mathcal{L}(\mathbf{d}|\boldsymbol{\theta})
\,
\pi(\boldsymbol{\theta})
\end{equation}
using the affine-invariant ensemble sampler implemented in \texttt{emcee} \citep{ForemanMackey2013}. This sampler is well suited to the present problem because the posterior distribution can be strongly anisotropic and non-Gaussian, especially for systems with poorly constrained inclination or weak photocentre motion.
Each system is sampled with 128 walkers, initialized near the Gaia catalogue solution for the orbital and astrometric parameters and from their respective priors for $M_1$, $M_2$, $\cos i$, and $\Omega$. The proposal distribution uses a mixture of Differential Evolution \citep{Nelson2014} and Differential Evolution Snooker \citep{terBraakVrugt2008} moves, in an 80\%/20\% ratio, which we find improves sampling efficiency for the strongly correlated posteriors typical of this problem. Convergence is assessed from the integrated autocorrelation time, and chains are run and, where necessary, reprocessed with longer lengths, modified proposals, or broader initialization, until this criterion and a target acceptance fraction are both satisfied. Systems that remain poorly convergent after this multi-pass procedure are flagged and excluded from the quantitative population-level conclusions of Section~\ref{sec:results}. The full sampler configuration including walker initialization, proposal mixture, pilot-run tuning, convergence criteria, burn-in and thinning, and the multi-pass strategy are given in Appendix~\ref{app:mcmc_details}.

\subsection{Posterior summaries and validation}
\label{subsec:posterior_summaries}
For each system, we report the 16th, 50th, and 84th percentiles of the posterior distributions. The median is used as the point estimate, while the 16th--84th percentile interval is quoted as the credible interval. In addition to the sampled parameters, we compute posterior distributions for derived quantities such as the photocentre semi-major axis $a_0$, the primary barycentric semi-major axis $a_1$, the mass ratio
\begin{equation}
q= \frac{M_2}{M_1},
\end{equation}
and the total mass $M_{\rm tot}=M_1+M_2$. The validation sample is used to compare the inferred companion mass with the Gaia catalogue value for systems in which both component masses are reported. This comparison provides a direct test of whether the adopted likelihood, priors, and covariance treatment recover known catalogue masses. The inference sample is then used to study the companion mass distribution of systems for which the secondary mass is not available as an input quantity.
\begin{figure*}
\includegraphics[width=0.45\textwidth]{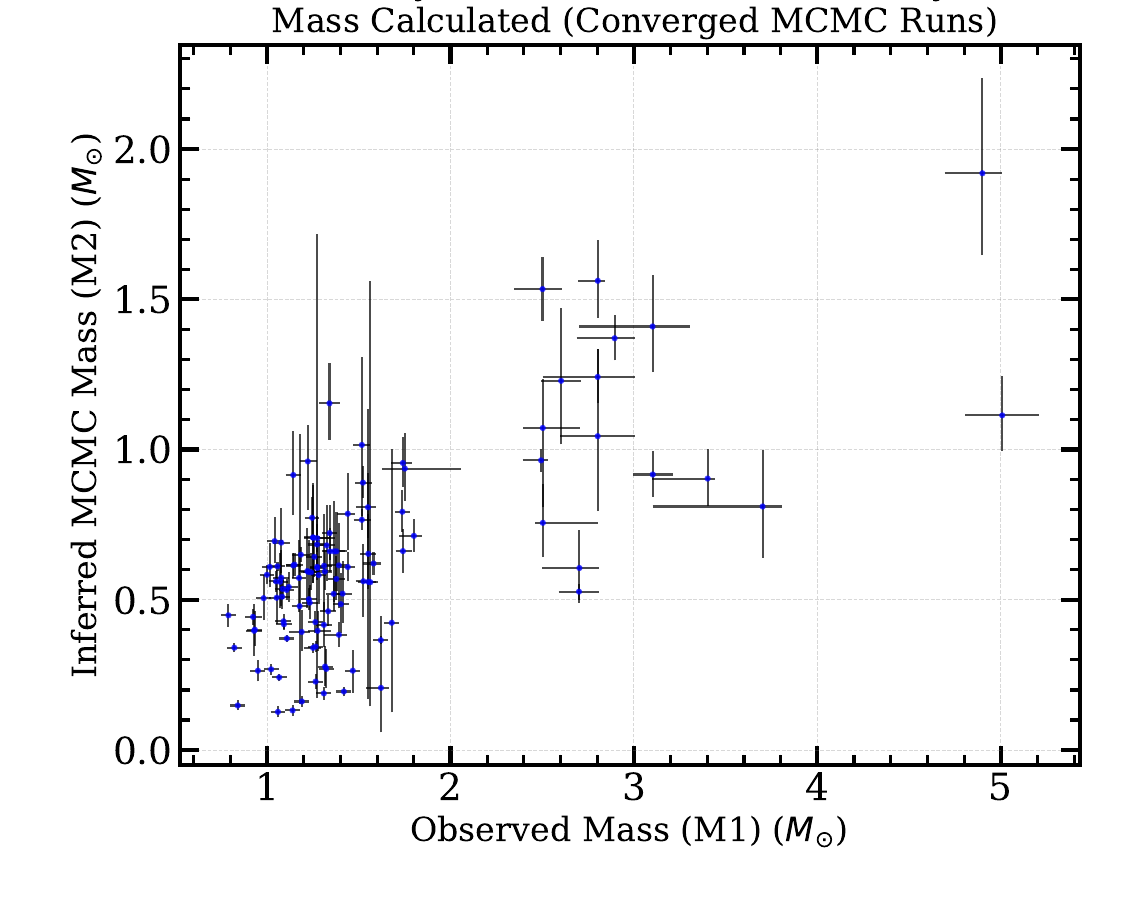}
\includegraphics[width=0.45\textwidth]{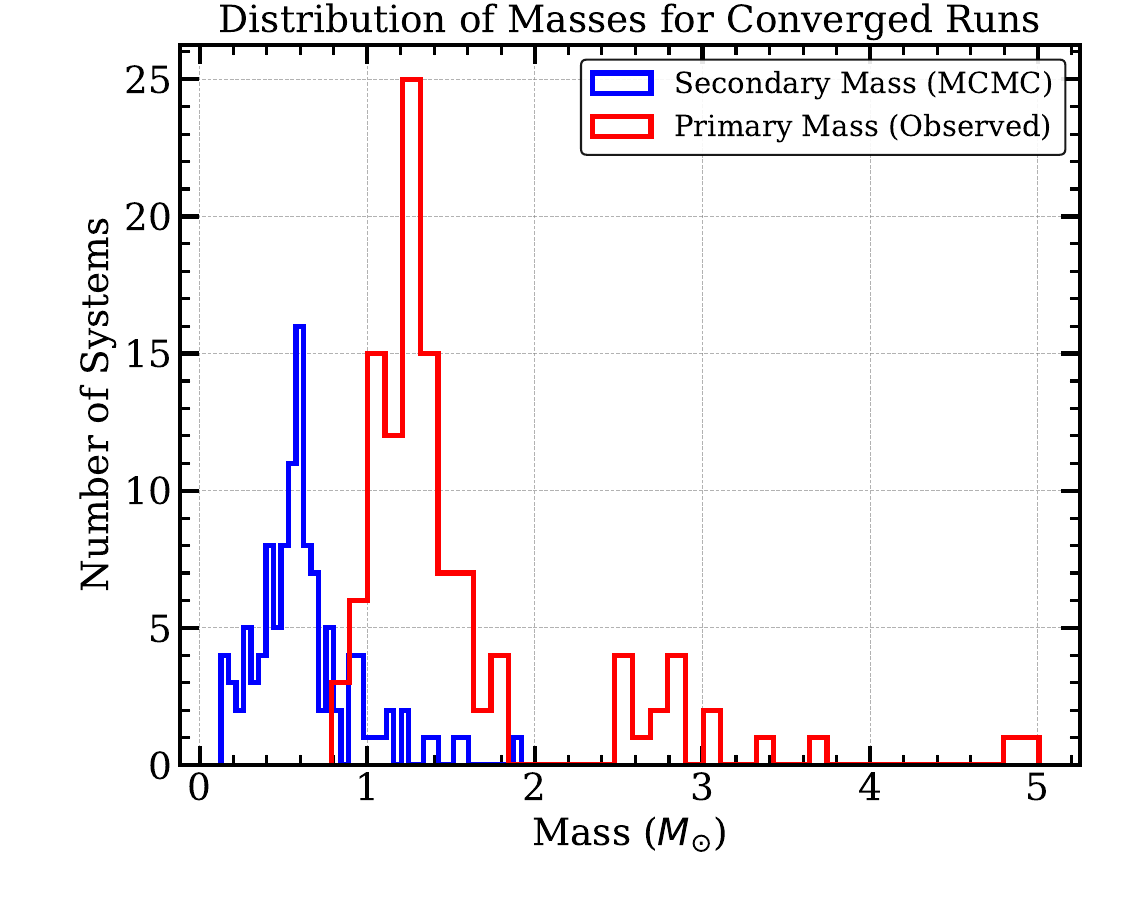}
\caption{ {\it Left panel:} We present 113 inferred companion mass $M_2$ as a function of primary mass $M_1$ (observed or catalogue mass) for the well converged inference sample. Points show posterior medians and error bars indicate the 16th--84th percentile credible intervals. Most systems are consistent with ordinary stellar companions, while a small high-mass tail contains systems that may merit further inspection as possible compact companion candidates. {\it Right panel:} Mass distribution of the converged inference sample. Primary component is shown in red and the inferred secondary component in blue.}
    \label{fig:m2_vs_m1}
\end{figure*}

\section{Results}
\label{sec:results}
We applied the Bayesian inference framework described in Section~\ref{sec:methods} to the  $1035$ Gaia DR3 \texttt{AstroSpectroSB1} systems identified in Section~\ref{sec:binarymasses}. The analysis was carried out in two stages. First, we considered a validation sample of $760$ systems for which tables \texttt{binary\_masses} $\cap$ \texttt{astrophysical\_parameters} reports both component masses. This sample is used to test whether the inference procedure can recover the catalogue value of the secondary mass when the same Gaia orbital information and covariance structure are used. Second, we then apply the method to the remaining  $275$ systems, for which only the primary mass is available. Therefore the secondary mass is an inferred quantity. The results presented in the main text focus on this inference sample, while the validation sample is discussed separately in Appendix~\ref{app:validation}.

\subsection{Convergence and sample quality}
\label{subsec:convergence_results}
Not all systems yield posterior samples of equal quality. This is expected because Gaia \texttt{AstroSpectroSB1} systems span a wide range of orbital periods, inclinations, parallaxes, eccentricities, and signal-to-noise ratios. We therefore classify the MCMC runs according to their sampling quality before interpreting the inferred masses. We use three quality categories. Category A contains systems for which the chains are well mixed. These systems form the primary science sample. Category B contains systems with acceptable acceptance fractions but autocorrelation times too long to satisfy the strict convergence criterion. These systems are useful for qualitative checks, but are not used for precision population-level conclusions. Category C contains systems with poor mixing, low acceptance fractions, or posterior structure dominated by boundaries or degeneracies. These systems are flagged and excluded from the main scientific interpretation. We summarize these in Table~\ref{tab:convergence_breakdown}, see also Appendix~\ref{app:mcmc_details} for further details. For the inference sample, $113$ of the $275$ systems satisfy the Category A criterion.  These well-converged systems are used in the results below. The remaining systems require either longer chains, alternative parameterizations, or additional external information before their companion-mass posteriors can be interpreted robustly. This quality cut is important because the inferred secondary mass can be especially sensitive to inclination for nearly face-on systems.

\begin{table}
  \centering
  \caption{MCMC convergence and acceptance-fraction breakdown for the  validation ($N=760$) and inference ($N=275$) samples  (Section~\ref{subsec:convergence_results}), mapped onto the Category  A/B/C classification of that section. ``Converged'' denotes  $N_{\rm chain}/\tau_{\rm max}\geq50$.  ``$f_{\rm acc}$ in range'' denotes $0.20<f_{\rm acc}<0.50$. Systems with no available full correlation vector (last row) were excluded before MCMC sampling and are not classified as A, B, or C. See Appendix~\ref{app:mcmc_details} for details.}
  \label{tab:convergence_breakdown}
  \footnotesize
  \begin{tabular}{llrr}
    \hline
    Cat. & Criterion & Validation & Inference \\
    \hline
    A & Converged, $f_{\rm acc}\in(0.20,0.50)$      & 345 & 113 \\
    B & Not converged, $f_{\rm acc}>0.20$            & 169 & 59  \\
    C & Not converged, $f_{\rm acc}\leq0.20$          & 231 & 97  \\
    C & Converged, $f_{\rm acc}\notin(0.20,0.50)$     & 13  & 5   \\
    -- & No full correlation vector                        & 2   & 1   \\
    \hline
     & Total                                          & 760 & 275 \\
    \hline
  \end{tabular}
\end{table}

\begin{figure*}
    \centering
    \includegraphics[width=0.48\textwidth]{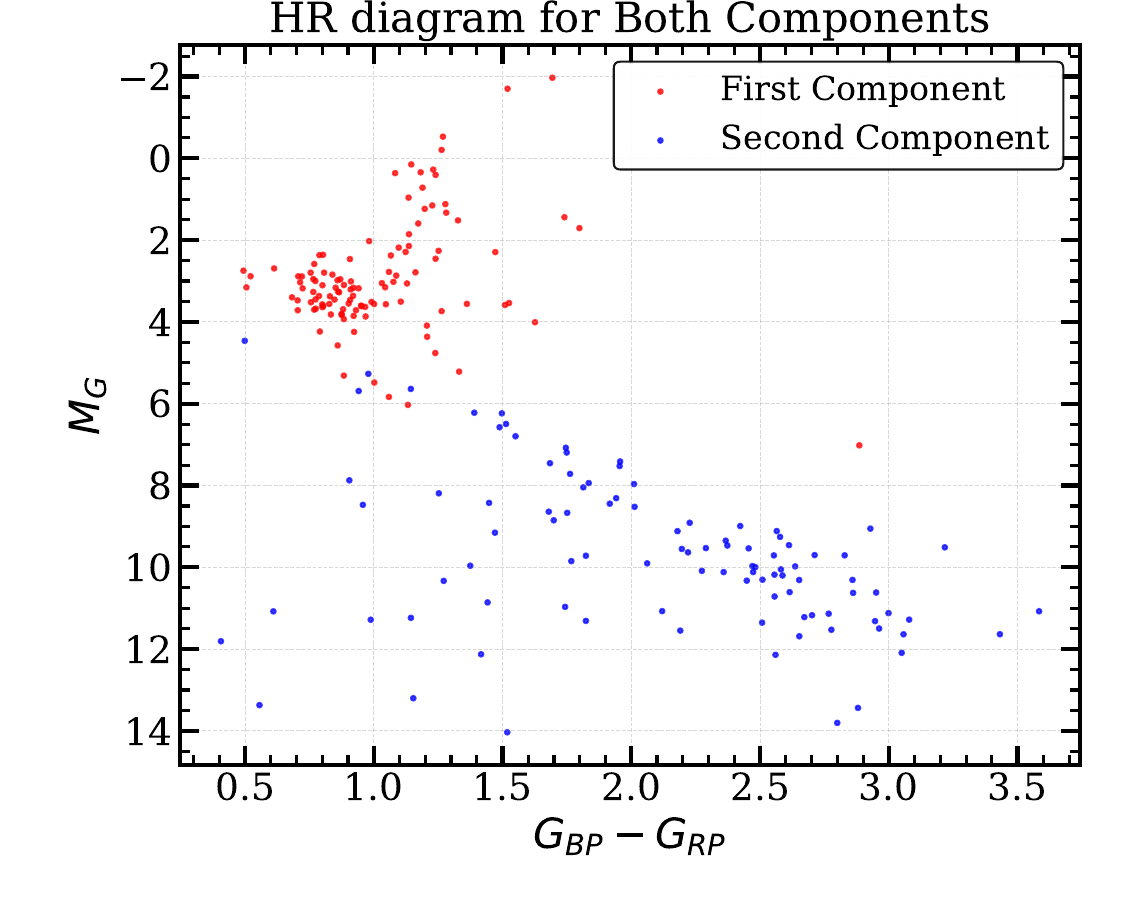}
    \includegraphics[width=0.48\textwidth]{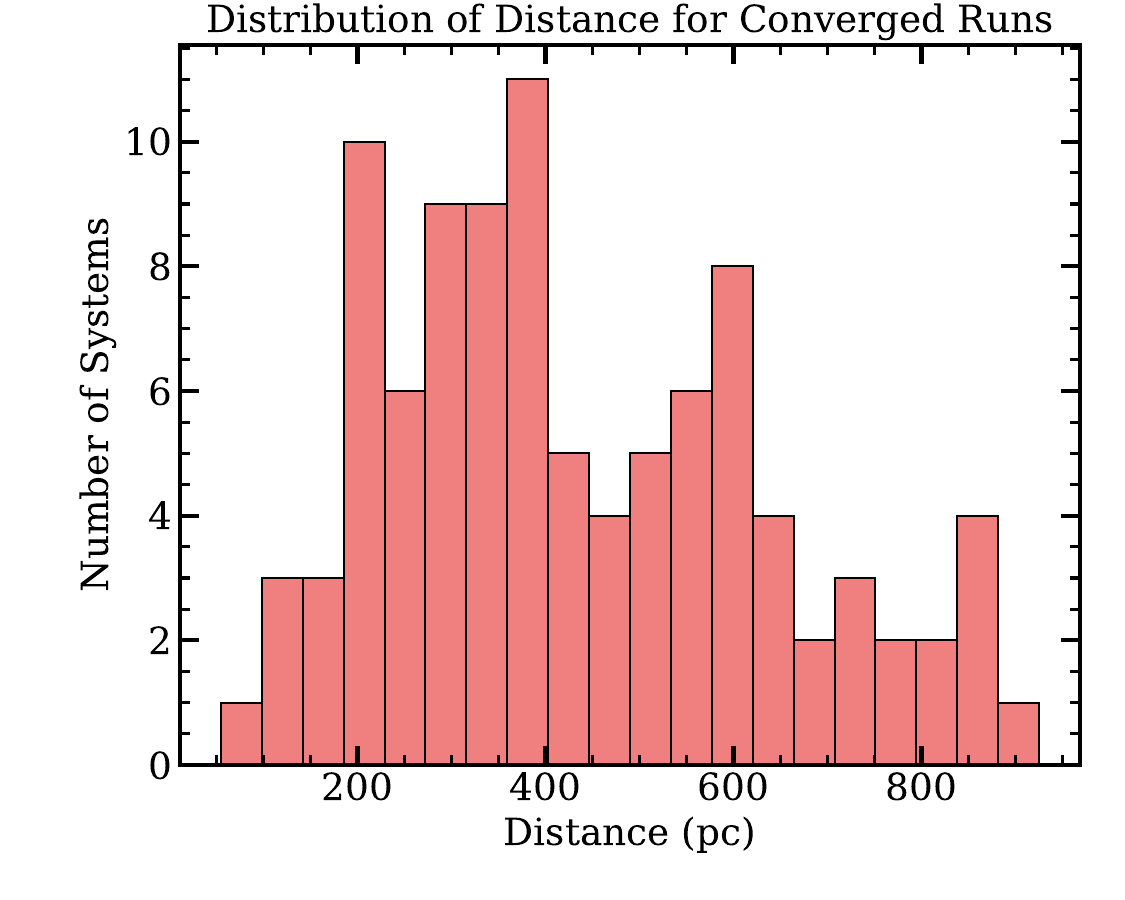}
    \caption{{\it Left panel:} Colour--magnitude diagram for the converged inference sample. The primary components are shown in red and the inferred secondary components in blue. The primary stars occupy the brighter locus, while the secondaries are shifted toward fainter magnitudes and redder colours, as expected for a population dominated by lower-mass stellar companions. {\it Right panel:} Distribution of Gaia parallax-based distance for the converged     inference sample.}
    \label{fig:cmd_components}
\end{figure*}

\subsection{Inferred companion masses}
\label{subsec:companion_masses}
The main inferred quantity is the secondary mass $M_2$. For each converged system, we quote the posterior median as the point estimate and the 16th--84th percentile interval as the credible range. Figure~\ref{fig:m2_vs_m1} shows the inferred companion mass as a function of the Gaia-provided primary mass for the converged inference sample (left panel), together with the marginal distribution of the inferred $M_2$ values across the sample (right panel). The inferred companion-mass distribution is dominated by ordinary stellar secondaries. Most systems lie below $M_2\simeq 1\,M_\odot$, as expected for a sample of unresolved binaries in which the secondary is generally fainter than the primary. A smaller number of systems extend toward larger inferred secondary masses. These high-mass systems are of particular interest because they may include massive white dwarfs or other compact-object candidates, and this is reflected in the marginal $M_2$ distribution in the right panel of Figure~\ref{fig:m2_vs_m1}, which is concentrated at lower masses with a sparser tail extending to higher values. However, such systems should not be classified on mass alone. Their interpretation requires careful assessment of convergence, inclination, flux-ratio assumptions, Gaia quality indicators, and possible contamination from luminous companions. We further note that the prior on $M_2$ is uniform on $[0.1,3.0]\,M_\odot$ (Section~\ref{sec:methods}); a posterior that accumulates near the upper edge of this range cannot be distinguished from a genuine physical cutoff on the basis of the posterior alone, and high-mass systems should therefore be checked individually for boundary pile-up before being discussed as compact-object candidates. The error bars in Figure~\ref{fig:m2_vs_m1} illustrate that the precision of $M_2$ varies significantly across the sample. Systems with favourable inclinations and well-constrained photocentre orbits have compact posteriors. In contrast, systems closer to face-on geometries can have broad or asymmetric posteriors, because the spectroscopic constraint depends strongly on $\sin i$. This behaviour is a generic feature of companion-mass inference in astrometric-spectroscopic binaries.

\subsection{Colour--magnitude diagram and distance distribution}
\label{subsec:cmd_results}
Figure~\ref{fig:cmd_components} shows the colour--magnitude diagram for the converged systems (left panel), together with the distribution of Gaia parallax-based distances for the same sample (right panel), included as a check that the inferred companion masses are not dominated by systems at a particular distance range where the adopted mass-luminosity mapping or the astrometric precision would be less reliable. The primary components occupy the brighter locus expected for the luminous stars used to define the Gaia \texttt{AstroSpectroSB1} solutions. The inferred secondary components are shifted toward fainter absolute magnitudes and generally redder colours. This behaviour is qualitatively consistent with a sample dominated by lower-mass stellar companions.

The secondary properties shown in this figure depend on the assumed mapping between mass, luminosity, and colour, and are therefore model dependent. Nevertheless, the fact that the secondaries occupy the expected fainter locus supports the broad physical plausibility of the inferred companion-mass distribution.

\subsection{Representative posterior structure}
\label{subsec:posterior_structure}
\begin{figure*}
    \centering
    \includegraphics[width=\textwidth]{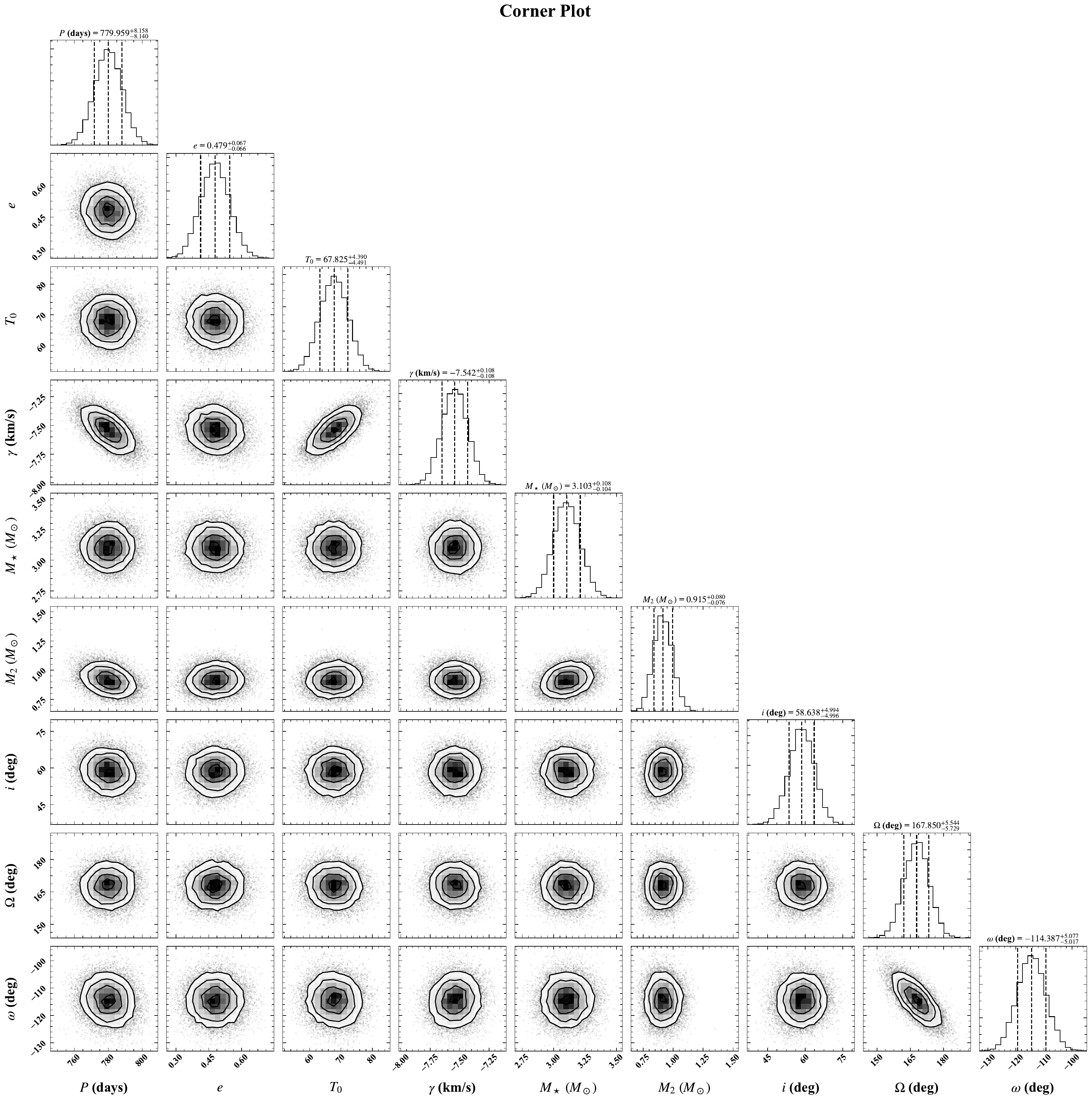}
    \caption{Representative corner plot for the system with Gaia DR3 source identifiers 547345929215245312 (primary) and 547345894855507200 (secondary). Diagonal panels show marginalized one-dimensional posteriors with the 16th, 50th, and 84th percentiles marked. Off-diagonal panels show joint posterior distributions. For this system, the primary mass is $M_1=3.1027\pm0.1081\,M_\odot$, and the inferred companion mass is $M_2=0.9151^{+0.0799}_{-0.0759}\,M_\odot$. }
    \label{fig:corner_example}
\end{figure*}

To illustrate the structure of the posterior distribution for an individual system, Figure~\ref{fig:corner_example} shows a corner plot for the binary with Gaia DR3 source identifiers 547345929215245312 (primary) and 547345894855507200 (secondary). In this system,
the primary mass is
\[
M_1 = 3.1027 \pm 0.1081\,M_\odot,
\]
while the inferred secondary mass is
\[
M_2 =
0.9151^{+0.0799}_{-0.0759}\,M_\odot.
\]
The marginalized posterior for $M_2$ is compact and well separated from the prior boundaries, indicating that the companion mass is constrained by the Gaia orbital information rather than by the imposed prior. The off-diagonal panels show the expected covariances among mass, inclination, and orbital geometry. Such corner plots are useful because they reveal whether the inferred companion mass is robust or whether it is driven by degeneracies. In particular, strong correlations between $M_2$ and $\cos i$ are expected in systems where the inclination is weakly constrained. Systems with broad, multimodal, or boundary-dominated posteriors are therefore not included in the Category A converged sample.

\subsection{Validation against systems with catalogue secondary masses}
\label{subsec:validation_summary}
The validation sample provides an important check on the inference framework. For these systems, Gaia reports estimates of both component masses, allowing the inferred secondary mass to be compared with a catalogue value. The detailed validation plots are presented in Appendix~\ref{app:validation}. In brief, the converged validation systems show broad consistency between the inferred and catalogue secondary masses. The residuals $\Delta M_2 = M_{2,\rm cat}-M_{2,\rm MCMC}$,  are distributed approximately symmetrically around zero, with no evidence for a strong global bias. The validation also confirms the usefulness of the convergence classification. Category A systems show the tightest agreement with the catalogue masses, while lower-quality chains exhibit larger scatter and more frequent outliers. This supports the use of the Category A cut when interpreting the companion-mass distribution of the inference sample.

\subsection{Summary of main results}
\label{subsec:results_summary}
The main conclusions from the inference sample are as follows. First, the MCMC framework successfully reproduces the Gaia astrometric, spectroscopic, and orbital observables for the well-converged systems. Second, the inferred companion-mass distribution is dominated by ordinary stellar companions, with a smaller high-mass tail that may contain interesting systems for follow-up. Third, the precision of the inferred secondary mass depends strongly on the orbital geometry, especially inclination. Finally, the validation sample indicates that, when the chains are well converged, the inferred companion masses are broadly consistent with catalogue values. These results show that Gaia DR3 \texttt{AstroSpectroSB1} solutions contain sufficient information for statistically meaningful companion-mass inference, provided that the full covariance structure is retained and systems with poor MCMC convergence are treated cautiously.


\section{Conclusions}
\label{sec:conclusion}
We have developed a Bayesian framework to infer companion masses in unresolved Gaia DR3 binaries with \texttt{AstroSpectroSB1} solutions. These systems are particularly valuable because Gaia provides both astrometric and spectroscopic orbital information, allowing the sky-plane photocentre motion and line-of-sight velocity information to be modelled within a common physical binary orbit. Rather than treating the Gaia orbital parameters as independent measurements, our method uses the full Gaia covariance matrix to propagate the correlated uncertainties in the astrometric, spectroscopic, and orbital parameters into posterior distributions for the physical binary quantities.

Starting from the full set of Gaia DR3 \texttt{AstroSpectroSB1} solutions, we restricted the analysis to hosts that also appear in an independently constructed, wide-separation binary catalogue first established in~\citep{million_binaries} for EDR3, here extended  to DR3. We adopted a Gaussian prior on the primary mass from Gaia's \texttt{astrophysical\_parameters} table, independent of the \textsc{AstroSpectroSB1} orbital solution itself, and then divided the resulting sample according to the availability of an independent secondary-mass estimate in Gaia's \texttt{binary\_masses} table. A validation sample of systems with both an \texttt{astrophysical\_parameters} \,\texttt{mass\_flame} and a \texttt{binary\_masses} secondary mass was used to test the inference procedure, while the remaining systems, for which the secondary mass is not available as an input quantity, were used to infer companion-mass posteriors. 

The forward model maps the sampled physical parameters to the Gaia-observed quantities, including the five astrometric parameters, the Thiele--Innes coefficients, the spectroscopic coefficients, the eccentricity, orbital period, time of periastron passage, and systemic velocity.
For the well-converged systems, the MCMC posterior samples reproduce the Gaia catalogue observables and recover the astrometric and spectroscopic orbital coefficients. This demonstrates that the same physical binary model can consistently describe both the photocentre orbit and the spectroscopic motion of the luminous primary. The inferred companion-mass distribution is dominated by ordinary stellar companions, as expected for a sample of unresolved binaries, and the inferred secondaries occupy the fainter, redder locus expected of lower-mass companions in the colour--magnitude diagram. The accompanying distance distribution shows no evidence that this result is driven by systems in a particular distance range (Section~\ref{subsec:cmd_results}). Representative posterior structure for individual systems shows that, when the Gaia orbital information is informative, the inferred companion mass is well separated from the prior boundaries and driven by the data rather than by the prior, though systems closer to face-on geometries can yield broad or asymmetric posteriors because the spectroscopic constraint depends strongly on $\sin i$ (Section~\ref{subsec:posterior_structure}). A smaller number of systems extend to higher inferred companion masses. These objects are potentially interesting as massive white-dwarf or neutron star candidates, but their interpretation requires caution because the inferred mass can be sensitive to inclination, flux-ratio assumptions, covariance quality, and possible systematics in the Gaia solution (Section~\ref{subsec:companion_masses}) before being reported as compact-object candidates.

The validation sample provides an important check on the reliability of the method. For the converged validation systems, the residuals between the inferred and catalogue secondary masses are distributed approximately symmetrically about zero, with no evidence for a strong global bias. The comparison also confirms the usefulness of the convergence classification: Category A systems show the tightest agreement with the catalogue masses, while lower-quality chains exhibit larger scatter and more frequent outliers. We therefore distinguish between well-converged systems, partially converged systems, and systems whose posteriors should not be interpreted without further analysis, and apply the same Category A cut when discussing the companion-mass distribution of the inference sample (Appendix~\ref{app:validation}).

Several improvements can strengthen the analysis in future work. First, the treatment of the flux ratio can be made more self-consistent by allowing the secondary light contribution to vary with the inferred companion mass using stellar-evolution models. Second, Gaia quality indicators and additional photometric or spectroscopic information can be incorporated to identify systems with problematic astrometric solutions or contaminating light. Third, the high-mass tail of the inferred companion-mass distribution can be ranked using posterior probabilities such as $P(M_2>1.25\,M_\odot)$ and $P(M_2>1.38\,M_\odot)$, producing a prioritized list for follow-up observations. A further extension, ranking candidates by $P(M_2>3\,M_\odot)$ to flag possible black-hole companions, would require relaxing the current $3.0\,M_\odot$ prior ceiling on $M_2$ (Section~\ref{sec:methods}) in a dedicated follow-up analysis, since posterior mass is by construction excluded from accumulating at or beyond that boundary under the present framework. 

Finally, the forthcoming Gaia Data Release 4 (DR4), expected on 2 December 2026,\footnote{\url{https://www.cosmos.esa.int/web/gaia/dr4}} will bear directly on several of these points. The flux ratio assumed here to be tested directly against the data rather than only modeled through stellar-evolution tracks, its non-single star processing extends to orbital models for systems with more than two bodies, offering a possible native alternative to the two-step wide-binary cross-match used in this paper (Section~\ref{sec:crossmatch}), and its roughly 66-month time baseline, nearly double that underlying Gaia DR3, is expected to improve orbital solutions for the longest-period systems most relevant to black-hole candidates. Re-applying the framework developed here to the DR4 successor of the \textsc{AstroSpectroSB1} solution type, once its exact naming and schema are established, is accordingly a natural next step.
The main result of this work is therefore methodological and statistical: Gaia DR3 \texttt{AstroSpectroSB1} solutions contain sufficient information for companion-mass inference when the full covariance structure is retained and the orbital solution is interpreted through a physical forward model. The framework developed here provides a scalable route to characterize large samples of unresolved binaries and to identify systems that may merit follow-up as candidate massive white-dwarf, neutron-star, or black-hole companions.

\section*{Data Availability}
The Gaia DR3 data used in this work are publicly available from the Gaia Archive (\url{https://gea.esac.esa.int/archive/}). The
wide-binary catalogue, MCMC posterior samples, and analysis code developed for this paper will be made available in a public repository upon publication, or before then upon reasonable request. The main results of the paper, including Gaia source-ids and other attributes, can be found in the \href{https://github.com/4prasid/Gaia_inferred_binary_masses.git}{GitHub repository}.

\section*{Acknowledgments}
R.G. acknowledges support from grant number  ANRF/ECRG/2025/000418/PMS. This work has made use of data from the European Space Agency (ESA) mission \emph{Gaia} (\url{https://www.cosmos.esa.int/gaia}), processed by the \emph{Gaia} Data Processing and Analysis Consortium (DPAC,\url{https://www.cosmos.esa.int/web/gaia/dpac/consortium}). Funding for the DPAC has been provided by national institutions, in particular the institutions participating in the \emph{Gaia} Multilateral Agreement \citep{Gaia:2016, Gaiadr3:2023}. This work made use of \texttt{emcee} \citep{ForemanMackey2013}, together with standard scientific Python packages including \texttt{NumPy}, \texttt{SciPy}, \texttt{Astropy}, \texttt{Astroquery}, and \texttt{Matplotlib}. We acknowledge National Supercomputing Mission (NSM) for providing computing resources of ‘PARAM RUDRA’ at P. G. Senapathy Center For Computer Resources, Play Field Ave, Indian Institute Of Technology, Chennai, Tamil Nadu 600036, which is implemented by C-DAC and supported by the Ministry of Electronics and Information Technology (MeitY) and Department of Science and Technology (DST), Government of India. During preparation of this manuscript, the authors used Claude (Anthropic) to assist with formatting LaTeX tables. The authors reviewed and take full responsibility for all AI-assisted content, and all scientific analysis, interpretation, and conclusions are their own.

\bibliographystyle{mnras}
\bibliography{biblio}


\appendix
\section{MCMC implementation and convergence diagnostics}
\label{app:mcmc_details}
This appendix gives the full technical specification of the MCMC sampling procedure summarized in Section~\ref{subsec:mcmc}: walker initialization, proposal strategy, run tuning, convergence criteria, and the multi-pass sampling strategy used to handle the range of posterior complexity across the sample.

\subsection{Sampler configuration}
The posterior is sampled using the affine-invariant ensemble sampler implemented in \texttt{emcee} \citep{ForemanMackey2013}. Each system is run with $N_{\rm walkers}=128$ walkers, initialized close to the Gaia catalogue solution for the astrometric and orbital parameters. The primary mass is initialized from its Gaussian prior, while $M_2$, $\cos i$, and $\Omega$ are initialized from broad distributions.
The proposal strategy uses an 80\%/20\% mixture of Differential Evolution \citep{Nelson2014} and Differential Evolution Snooker \citep{terBraakVrugt2008} moves, implemented as \texttt{DEMove} and \texttt{DESnookerMove} in \texttt{emcee}. This combination improves sampling in posteriors with strong correlations, especially those involving $M_2$, $\cos i$, and the orbital coefficients.
A pilot run of 2000 steps is first used to estimate the acceptance fraction ($f_{\rm acc}$). We require
\begin{equation}
0.20<f_{\rm acc}<0.50.
\end{equation}
If the acceptance fraction lies outside this interval, the proposal scale is adjusted and the pilot run is repeated.

\subsection{Convergence criteria and quality flags}
Convergence is assessed using the integrated autocorrelation time $\tau$. For each parameter we estimate $\tau_i$, and define
\begin{equation}
\tau_{\rm max}=\max_i\tau_i.
\end{equation}
A chain is considered converged if
\begin{equation}
N_{\rm steps}\geq50\,\tau_{\rm max}.
\end{equation}
After convergence, the first $2\tau_{\rm max}$ steps are discarded as burn-in. The remaining samples are thinned by approximately $\tau_{\rm max}/2$ to obtain nearly independent posterior samples. Together with the acceptance-fraction criterion above, this convergence criterion defines the Category A, B, and C sampling-quality flags used to interpret the inferred companion masses in Section~\ref{subsec:convergence_results}: Category A systems satisfy both criteria and form the primary science sample; Category B systems have an acceptable acceptance fraction but insufficient chain length relative to $\tau_{\rm max}$; and Category C systems are poorly mixed, boundary-dominated, or dominated by strong parameter degeneracies. Only Category A systems are used for the main population-level conclusions.

\subsection{Multi-pass sampling strategy}
Because the posterior complexity varies substantially across the Gaia \texttt{AstroSpectroSB1} sample, the analysis is carried out in multiple passes. In the first pass, all systems are run in parallel with a fixed maximum number of steps. Systems that do not converge in this pass are rerun with longer chains. A final pass is reserved for problematic systems, for which broader initialization, modified proposal mixtures, or tempered warm-up runs are attempted.
Systems that remain poorly mixed after these attempts are not discarded from the catalogue, but they are flagged. Their posterior summaries are treated as a diagnostic instead of a robust companion-mass inference.

\section{Recovery of known companion masses}
\label{app:validation}
This appendix demonstrates that the inference procedure can recover known companion masses for systems in which Gaia reports estimates of both components. These systems form the validation sample and provide a direct test of the likelihood, covariance treatment, and MCMC implementation.

\subsection{Validation sample}
The validation sample consists of systems with \texttt{AstroSpectroSB1} solutions for which Gaia provides catalogue estimates of both the primary mass $M_1$ and the secondary mass $M_2$. In the validation exercise, the primary mass is used as a prior in the same way as for the inference sample, while the secondary mass is not used as an input to the MCMC. The inferred posterior distribution for $M_2$ is then compared with the Gaia catalogue value.
This test is intended to check whether the inference pipeline, when applied to Gaia orbital solutions and Gaia mass information, produces companion masses consistent with the catalogue values for systems where such values are available.

\subsection{Recovery of Gaia catalogue observables}
\label{subsec:observable_recovery}
\begin{figure*}
    \centering
    \includegraphics[width=\textwidth]
    {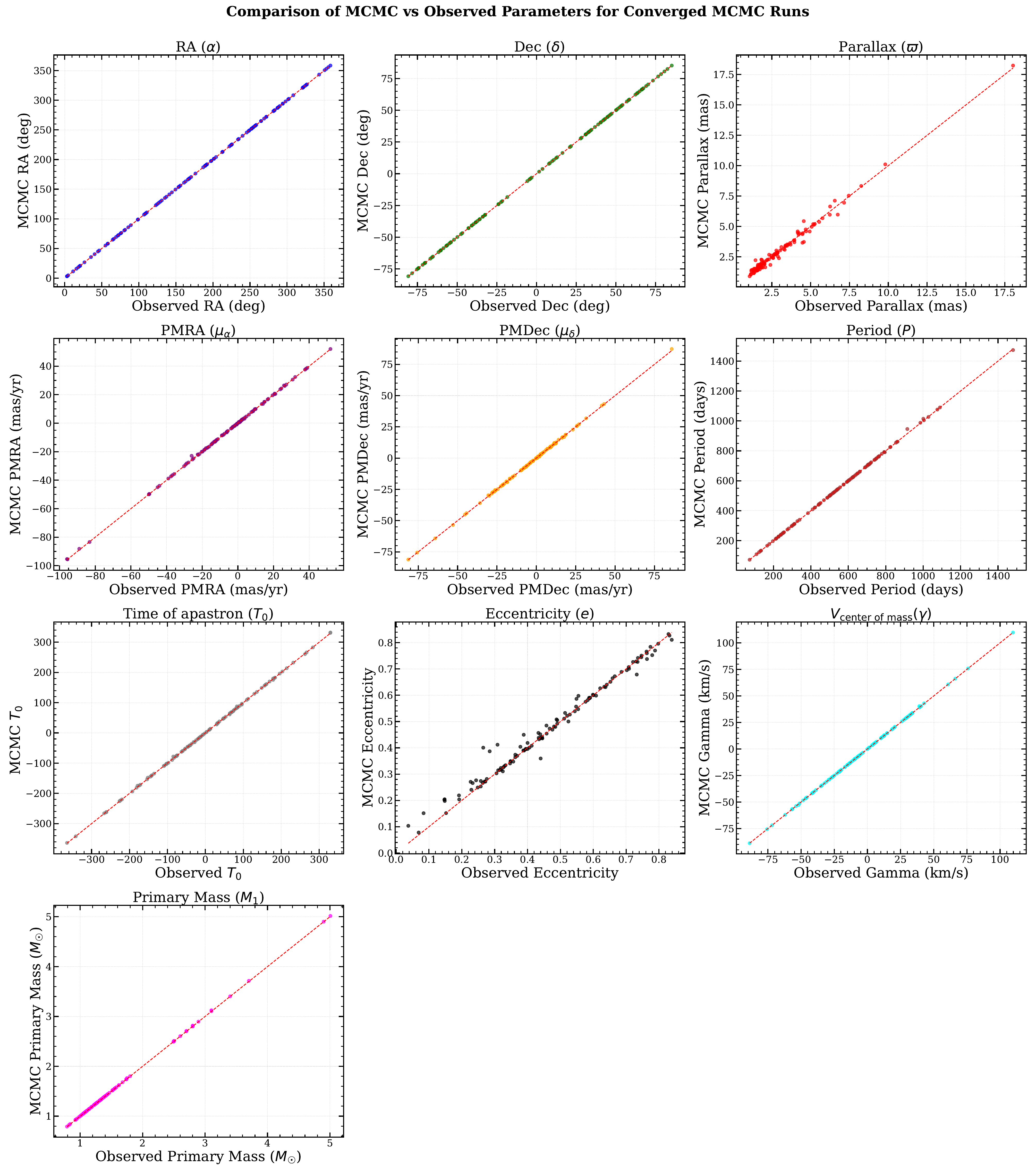}
    \caption{
    Comparison between MCMC-derived posterior medians and Gaia DR3 catalogue values for the validation sample. The panels show the sky position, parallax, proper motions, orbital period, time of periastron passage, eccentricity, systemic velocity, and catalogue primary mass, respectively. The dashed red line shows the one-to-one relation. The close agreement verifies that the likelihood and covariance implementation recover the Gaia observables used as input to the inference.
    }
    \label{fig:catalogue_recovery}
\end{figure*}

A basic requirement of the inference framework is that posterior samples reproduce the Gaia observables used in the likelihood. Figure~\ref{fig:catalogue_recovery} compares the posterior median values of the directly constrained quantities with the corresponding Gaia catalogue values for the validation sample. The plotted quantities include right ascension, declination, parallax, proper motion in both coordinates, orbital period, time of periastron passage, eccentricity, systemic velocity, and the primary mass.
The strong clustering of points around the one-to-one relation shows that the sampler correctly recovers the quantities that are directly constrained by the Gaia solution. This is an important sanity check on the likelihood implementation and covariance handling. The astrometric parameters and orbital period show particularly tight agreement with the catalogue values. The largest visible scatter occurs in eccentricity, which is expected: for nearly circular systems, small changes in the sampled parameters $e\cos\omega$ and $e\sin\omega$ can produce relatively large fractional changes in $e$, while leaving the observable orbit nearly unchanged.

\begin{figure*}
    \centering
    \includegraphics[width=\textwidth]
    {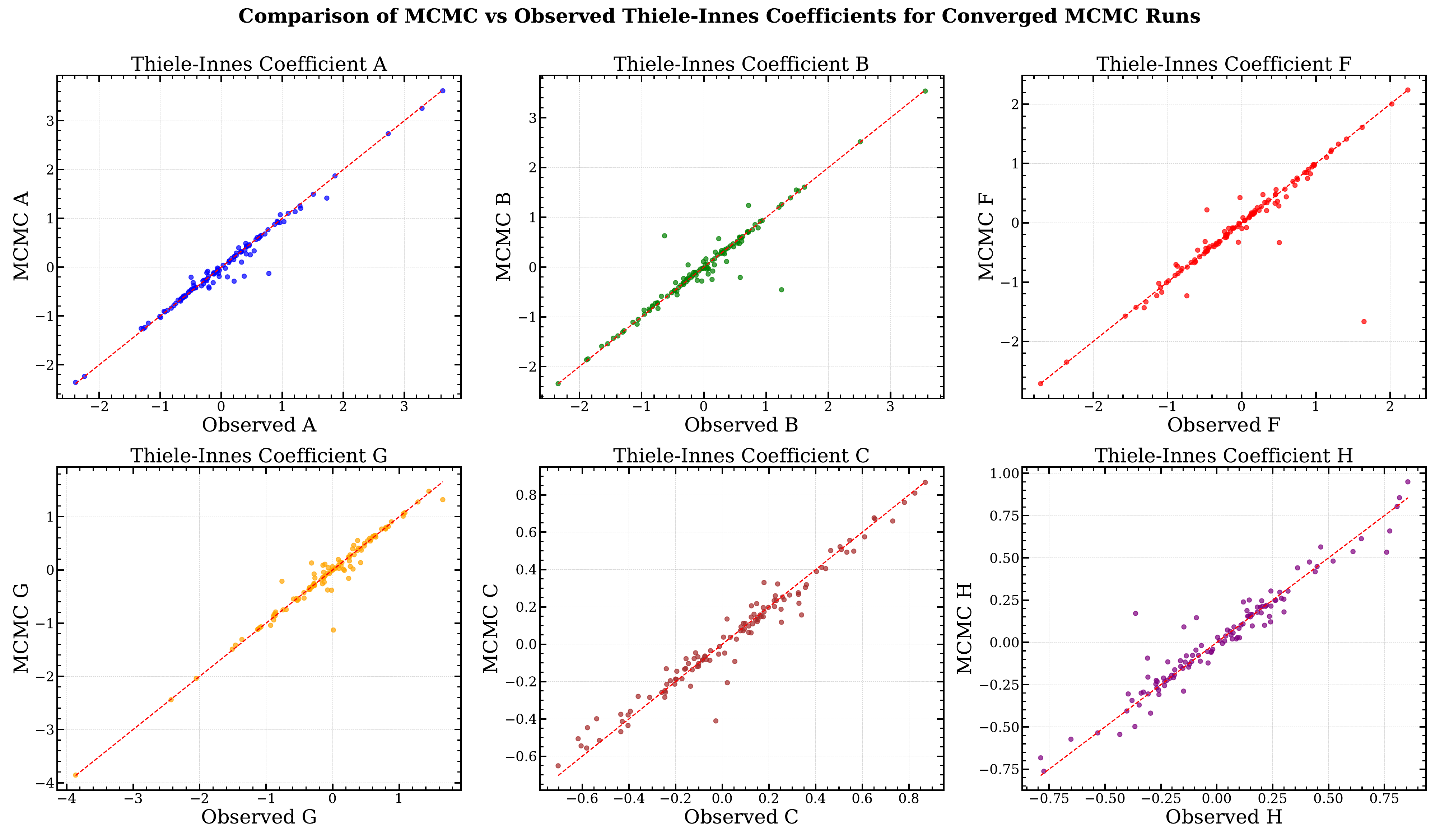}
    \caption{
    Comparison between Gaia DR3 orbital coefficients and the corresponding MCMC-derived posterior medians for the validation sample. The four astrometric Thiele--Innes coefficients $A,B,F,G$ and the two spectroscopic coefficients $C,H$ are shown separately. The dashed line marks equality. The recovery of both sets of coefficients confirms that the same physical binary model reproduces the astrometric and spectroscopic components of the Gaia \texttt{AstroSpectroSB1} solution.
    }
    \label{fig:ti_recovery}
\end{figure*}

In figure~\ref{fig:ti_recovery} we perform the same comparison for the orbital coefficients. The astrometric Thiele-Innes coefficients $A,B,F,G$. The spectroscopic coefficients $C,H$ are not sampled directly, instead, they are computed from the physical binary parameters at each posterior sample. The strong correlation between the observed and reconstructed values demonstrates that the forward model consistently maps the sampled masses and orbital geometry back to the Gaia astrometric and spectroscopic solution. This agreement is non-trivial. The same values of $M_1$, $M_2$, $i$, $\Omega$, and $\omega$ must simultaneously reproduce the sky-plane photocentre orbit and the spectroscopic orbit of the luminous primary. The recovery of both the astrometric and spectroscopic coefficients therefore shows that the model is internally consistent and that the inferred companion masses are not obtained by fitting the astrometric or spectroscopic parts in isolation.

\subsection{Mass-recovery for known systems}
For each validation system, we define the residual
\begin{equation}
\Delta M_2 = M_{2,\rm MCMC} - M_{2,\rm cat},
\label{eq:dm2}
\end{equation}
where $M_{2,\rm MCMC}$ is the posterior median inferred by the MCMC and $M_{2,\rm cat}$ is the Gaia catalogue value. We also define the normalized residual
\begin{equation}
z = \frac{ M_{2,\rm MCMC}-M_{2,\rm cat}}{\sigma_{M_2}},
\end{equation}
where $\sigma_{M_2}$ is taken to be the 68\% posterior uncertainty on the inferred companion mass. If the catalogue uncertainty on $M_{2,\rm cat}$ is available, we instead use
\begin{equation}
z=\frac{M_{2,\rm MCMC}-M_{2,\rm cat}}{\sqrt{\sigma^2_{M_{2,\rm MCMC}}+\sigma^2_{M_{2,\rm cat}}}}.
\label{eq:znorm}
\end{equation}
A useful summary of the validation performance is given by the median bias,
\begin{equation}
b_{M_2}={\rm median}\left(M_{2,\rm MCMC}-M_{2,\rm cat}\right),
\end{equation}
and the robust scatter,
\begin{equation}
\sigma_{\rm MAD}=1.4826\,{\rm median}\left[\left|\Delta M_2-{\rm median}(\Delta M_2)\right|\right].
\end{equation}

For a well-calibrated inference procedure, the residual distribution should be centred near zero, and the normalized residuals should be broadly consistent with a unit-width distribution for the well-converged systems.

\subsection{Comparison of inferred and catalogue masses}
\begin{figure*}
    \centering
 \includegraphics[width= 0.45\textwidth]{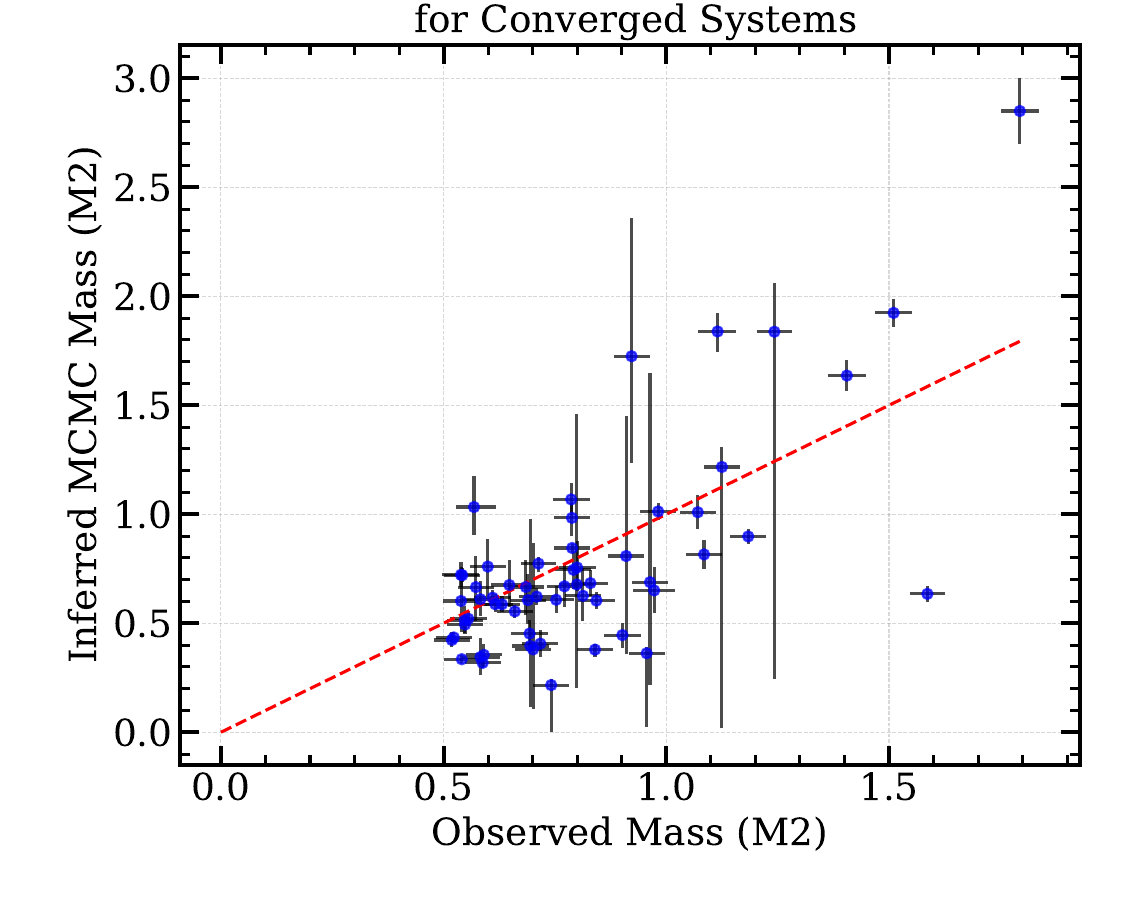} \includegraphics[width=0.45\textwidth]{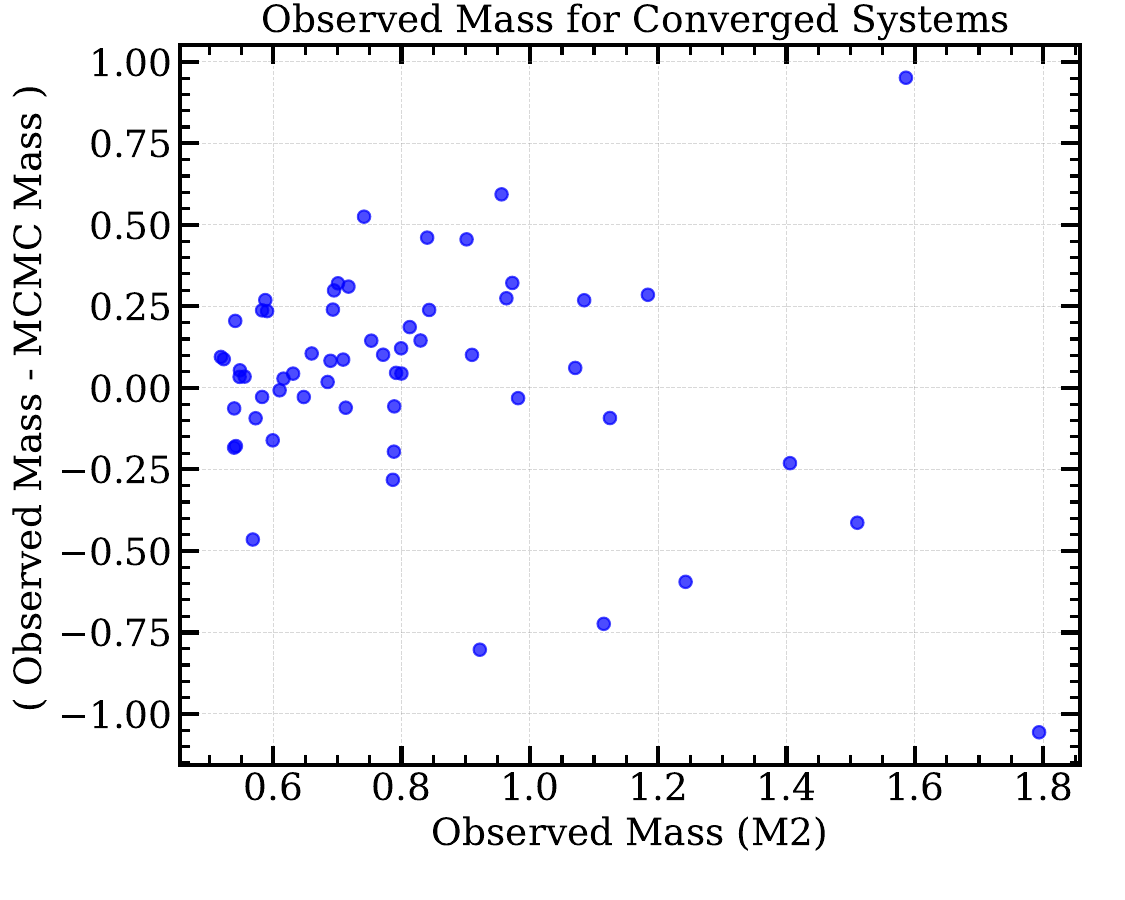}
    \caption{{\it Left panel:}
    Recovery of known companion masses in the validation sample. The horizontal axis shows the Gaia catalogue companion mass, $M_{2,\rm cat}$, while the vertical axis shows the posterior median inferred by the MCMC, $M_{2,\rm MCMC}$. Error bars show the 16th--84th percentile posterior interval. The dashed line denotes  equality. Well-converged systems should cluster around the one-to-one relation. {\it Right panel:} Residual mass for the validation sample. The residual is  defined as $\Delta M_2=M_{2,\rm MCMC}-M_{2,\rm cat}$. A distribution centred near zero indicates that the inference procedure does not introduce a strong global bias in the companion mass for the well-converged systems.}
    \label{fig:validation_m2_recovery}
\end{figure*}

In Figure~\ref{fig:validation_m2_recovery} left panel, we compare the inferred companion mass with the Gaia catalogue companion mass for the validation sample. Each point represents one binary system. The horizontal axis shows the Gaia catalogue value $M_{2,\rm cat}$, while the vertical axis shows the MCMC posterior median $M_{2,\rm MCMC}$. Error bars indicate the 16th--84th percentile credible interval of the inferred posterior. The dashed line shows the one-to-one relation.
For well-converged systems, the points cluster around the one-to-one line. This demonstrates that the MCMC inference recovers the catalogue companion masses when the Gaia orbital solution provides sufficient information and the posterior is well mixed. The scatter about the one-to-one relation is expected because the inference uses the full covariance matrix and propagates uncertainties in the astrometric, spectroscopic, and orbital parameters.

\subsection{Residual distribution}
Figure~\ref{fig:validation_m2_recovery} right panel shows the distribution of the residual $\Delta M_2$ Eq.~\eqref{eq:dm2} for the validation sample. The residuals for well-converged systems are approximately centred around zero, indicating no strong global bias in the recovered companion masses. Systems with broad or asymmetric posteriors contribute to the tails of the distribution and are typically associated with weaker inclination constraints or poorer MCMC convergence.

A complementary diagnostic is the normalized residual $z$ of Eq.~\eqref{eq:znorm}. If the uncertainties are well calibrated, the distribution of $z$ should be approximately centred at zero with width of order unity. Deviations from this behaviour indicate either underestimated uncertainties, catalogue-level mass systematics, or systems whose posteriors are not fully converged.

\subsection{Mass-dependent recovery limitations}
\label{subsec:mass_dependent_limits}
Beyond convergence quality, two effects specific to high companion mass can independently degrade recovery in the validation sample. The first is a prior-boundary effect rather than a sampling problem: because $M_2$ carries a uniform prior on $[0.1,3.0]\,M_\odot$ (Section~\ref{sec:methods}), the posterior for any validation system with $M_{2,\rm cat}$ close to this ceiling is truncated by the prior itself rather than shaped purely by the likelihood. Probability mass that the likelihood would place above $3.0\,M_\odot$ is instead redistributed within the allowed range, which systematically pulls the posterior median below the true value. This predicts a specific signature: $\Delta M_2$ should trend negative as $M_{2,\rm cat}$ approaches the prior boundary, independent of chain convergence.

The second is astrophysical, and follows directly from the
photocentre-offset relation of Section~\ref{sec:methods},
\[
a_0=\varpi\,a_{\rm rel}\left[\frac{M_2}{M_1+M_2}-f\right],
\]
where $f$ is the secondary's fractional light contribution. For an ordinary faint low-mass companion, $f\to0$ and $a_0$ is a clean, monotonic function of $M_2$. A sufficiently massive secondary, however, is also more luminous, so $f$ approaches the mass fraction itself and $a_0\to0$ regardless of the true companion mass. Such a system carries almost no astrometric leverage on $M_2$ even with an otherwise well-constrained orbit, and would also sit in tension with the assumptions underlying the \textsc{AstroSpectroSB1} solution type itself. These two effects are, in principle, distinguishable: the prior-boundary effect predicts a one-sided, systematic bias in $\Delta M_2$ that grows toward $M_{2,\rm cat}\to3.0\,M_\odot$, while the flux-ratio degeneracy predicts increased scatter (and, for the most extreme cases, non-Gaussian or boundary-dominated posteriors that should already be excluded by the Category A cut). We recommend examining $\Delta M_2$ explicitly as a function of $M_{2,\rm cat}$ for the validation sample before attributing poor recovery at high mass to either effect, since both predict degraded performance in the same regime but would call for different remedies. Relaxing the prior ceiling in the first case, versus incorporating an evolving flux-ratio model in the second (Section~\ref{sec:conclusion}).

\subsection{Dependence on convergence quality}
The validation sample also demonstrates the importance of the MCMC quality classification. Category A systems show the closest agreement between inferred and catalogue companion masses. Category B systems generally remain broadly consistent but show larger scatter, reflecting longer autocorrelation times or incomplete exploration of the posterior. Category C systems frequently show broad, multimodal, or boundary-dominated posteriors and should not be used for quantitative mass recovery. This behaviour motivates the use of Category A systems as the primary science sample in the main text. The validation test therefore serves two purposes: it verifies that the forward model can recover known companion masses, and it justifies the convergence cuts used when interpreting the inferred companion-mass distribution.

\subsection{Interpretation of the validation}
The recovery of Gaia-reported companion masses should be interpreted carefully. The validation does not independently verify the astrophysical accuracy of the Gaia masses. It verifies the internal consistency of the MCMC procedure relative to Gaia's published orbital and component-mass information. In particular, any systematic uncertainty in the Gaia mass estimates, the assumed flux ratio, or the published orbital solution will propagate into both the validation comparison and the inference sample.
A concrete instance of this concerns systems with moderately larger companion masses, for which $M_{2,\rm MCMC}$ can come out close to a factor of two larger than $M_{2,\rm cat}$. Because $f$ is fixed rather than fitted jointly with the masses, and because $a_0$ depends on the difference between the mass fraction $M_2/(M_1+M_2)$ and $f$ (Section~\ref{sec:methods}), any error in the adopted $f$ propagates directly into $M_2$ and is amplified once the two terms become comparable in size. This is precisely the regime where a companion is bright enough for $f$ to matter but still too faint to yield an independently measured flux ratio, by the single-lined definition of an \textsc{AstroSpectroSB1} solution. An adopted $f$ that overestimates the true light contribution of the close companion, for example through contamination from the resolved photometry of a wide common-proper-motion companion (Section~\ref{sec:widecat}) rather than the close pair itself, would inflate the inferred mass fraction and hence $M_2$ by a similar factor. We view this fixed-flux-ratio assumption, rather than convergence quality, as the most likely explanation for the intermediate-mass discrepancies discussed in Section~\ref{subsec:mass_dependent_limits}, and recommend it be checked directly against Gaia's own flux information before the method is applied more broadly.

Nevertheless, the validation exercise is essential. It shows that, for well-converged systems, the Bayesian forward model and covariance treatment reproduce the known catalogue companion masses within the expected uncertainties. This provides confidence that the same framework can be applied to systems for which the secondary mass is not provided as an input quantity.

\end{document}